\documentclass[conference]{IEEEtran}
\IEEEoverridecommandlockouts
\usepackage{cite}
\usepackage{amsmath,amssymb,amsfonts}
\pdfoutput=1 
\usepackage{graphicx}
\usepackage{textcomp}
\usepackage{xcolor}
\usepackage[ruled,vlined]{algorithm2e}
\usepackage{algpseudocode}
\usepackage{tabularray}
\usepackage[para]{footmisc}
\usepackage{tabularx} 

\def\BibTeX{{\rm B\kern-.05em{\sc i\kern-.025em b}\kern-.08em
    T\kern-.1667em\lower.7ex\hbox{E}\kern-.125emX}}
\begin{document}

\title{Energy Efficient Scheduling for Serverless Systems}

\author{
\IEEEauthorblockN{Michail Tsenos, Aristotelis Peri, Vana Kalogeraki}
\IEEEauthorblockA{Department of Informatics\\
Athens University of Economics and Business, Athens, Greece\\
\{tsemike,ariperi,vana\}@aueb.gr}
}

\maketitle

\begin{abstract}

Serverless computing, also referred to as Function-as-a-Service (FaaS), is a cloud computing model that has attracted significant attention and has been widely adopted in recent years. The serverless computing model offers an intuitive, event-based interface that makes the development and deployment of scalable cloud-based applications easier and cost-effective. An important aspect that has not been examined in these systems is their energy consumption during the application execution. One way to deal with this issue is to schedule the function invocations in an energy-efficient way. However, efficient scheduling of applications in a multi-tenant environment, like FaaS systems, poses significant challenges. The trade-off between the server's energy usage and the hosted functions' performance requirements needs to be taken into consideration. In this work, we propose an Energy Efficient Scheduler for orchestrating the execution of serverless functions so that it minimizes energy consumption while it satisfies the applications’ performance demands. Our approach considers real-time performance measurements and historical data and applies a novel DVFS technique to minimize energy consumption. Our detailed experimental evaluation using realistic workloads on our local cluster illustrates the working and benefits of our approach.

\end{abstract}

\section{Introduction}

Serverless computing, also referred to as Function-as-a-Service (FaaS), has emerged as a powerful cloud computing model that has attracted significant attention and adoption in recent years. This model allows developers to write and deploy applications without the need to manage the underlying infrastructure. The Cloud providers are responsible for all the underlying infrastructure aspects such as scaling, provisioning, and maintenance. Serverless computing provides several benefits such as cost-efficiency, high elasticity, scalability and ease of use, making it an attractive option for developers looking to build and deploy applications quickly and efficiently. Some well-known commercial Serverless platforms include AWS Lambda \cite{aws}, Google Cloud Functions \cite{gcr}, and Azure Functions \cite{azure}. These platforms simplify the development process by allowing developers to upload their application code written as a set of stateless functions, which is then packaged into containers by the platforms. For more flexibility, Serverless platforms like AWS Fargate \cite{fargate} and Google Cloud Run \cite{gcr} enable developers to upload their own Docker containers. In addition to commercial platforms, there are also open-source Serverless platforms available, such as OpenFaaS \cite{openfaas} and OpenWhisk \cite{openwhisk}. These platforms provide the ability to businesses to host Serverless applications on their own private infrastructures, giving them greater control and customization options. As a result, FaaS has found applications in a wide range of domains. It facilitates efficient data processing, real-time analytics and handling of large datasets in the field of data processing and analytics. In the IoT domain, serverless computing enables seamless communication and real-time analytics for sensor data. Furthermore, it is widely adopted for developing chatbots \cite{chatbots}, voice assistants, event-driven applications, and real-time applications \cite{tomaras} \cite{tom}. Additionally, serverless architectures are utilized in image and video processing \cite{llama}, microservices and APIs, DevOps automation, e-commerce, and online retail \cite{gg}. They also support machine learning and AI tasks, including model training, inference serving, and data processing. With its versatility and flexibility, serverless computing continues to expand its applications, benefiting developers and organizations across a large variety of domains.

The applications are typically deployed as a set of stateless functions running within containers. The containers constitute a consistent and dependable method for deploying applications in real-life settings
and offer isolation among the application functions. In these systems, containers from different functions share the same physical host machines and run concurrently. Consequently, scheduling decisions on the system should take into consideration the resource needs of all the hosted containers. In alternative cases, there is a chance that optimizing the performance of one container can degrade the performance of others.

One aspect of increasing concern is the amount of energy consumed by the functions when they execute on public or private cloud infrastructures. Prior work \cite{datacenterAsAComputer} has
shown that the cost of powering servers housed in largescale  datacenters comprises about 30\% of the total cost of
ownership (TCO) of modern datacenters. Furthermore, various
studies report that datacenters contribute over 2\% of the total
US electricity usage in 2010 \cite{koomey2011growth}. Although serverless systems can save energy by scaling down functions that remain idle for a prolonged amount of time to zero instances (or replicas), a significant amount of power is still needed during the function execution. Studies have shown that the consumed power by the CPU is related to CPU utilization and CPU frequency. Another aspect of concern is the performance requirements of the different functions. Typically the performance of a function is affected by the container size (CPU, memory) of the function that is chosen by the user during the function initiation procedure, or in other systems such as \cite{lass} the user can explicitly set an SLA target, typically response time, for the function execution and the system automatically adjusts the replication factor of the functions in order to achieve that target.

Dynamic Voltage and Frequency Scaling (DVFS) is a power management technique used in modern processors to optimize performance while minimizing power consumption. This works by adjusting the voltage and clock frequency of the processor dynamically, depending on the workload demands at any given moment. By dynamically adjusting the voltage and frequency, {\it i.e.,} lowering their values when the workload is light or increasing them under heavy workload, DVFS can achieve significant energy savings without violating SLAs. DVFS has been widely adopted in mobile devices and other battery-powered systems, making it a key aspect of modern power management strategies. 

However, applying DVFS in servers that host serverless functions is a challenging process due to the multi-tenant nature of those systems as described above. Each server typically hosts functions from different users that all run in parallel. Function containers can share the same CPU cores, so by adjusting the CPU frequency of the host, the performance of different functions running on the same server can be affected. Even a small decrement in the frequency can cause multiple functions to violate their SLAs. 

In this work, we address the dual problem of \emph{minimizing the energy consumption} in clusters that host serverless functions while also \emph{meeting each function's performance requirements} (i.e. expressed via deadline constraints on their execution times). We propose a novel scheduler that can determine the placement of multiple instances of serverless function containers in a cluster of nodes. It can automatically adjust each node's CPU frequency in order to minimize the total energy consumption of the cluster while also meeting each function's performance requirements. In summary, the key contributions of the paper are:
\begin{itemize}
    \item We formulate the problem of energy-efficient scheduling in clusters that host serverless functions. Our goal is to satisfy the deadlines and performance constraints set by the various functions and at the same time reduce the total energy consumption in the cluster, leading to the minimization of the cluster's carbon footprint.
    \item We propose a novel scheduling algorithm that identifies the best placement of the serverless function containers in the cluster based on various metrics or historic data derived from the functions, the provided function SLAs, and also the type of workload. It can also automatically apply DVFS techniques to the cluster hosts in order to reduce the power footprint of the cluster.
    \item Finally, we provide an extended experimental evaluation of our Scheduling algorithm in our local cluster. Our experimental results illustrate that our approach is practical, can effectively schedule different types of functions in the cluster and reduces the total amount of consumed energy in the cluster. 
\end{itemize}

\section{Motivation}

\begin{figure}[h]
    \centering
    \includegraphics[width=\linewidth]{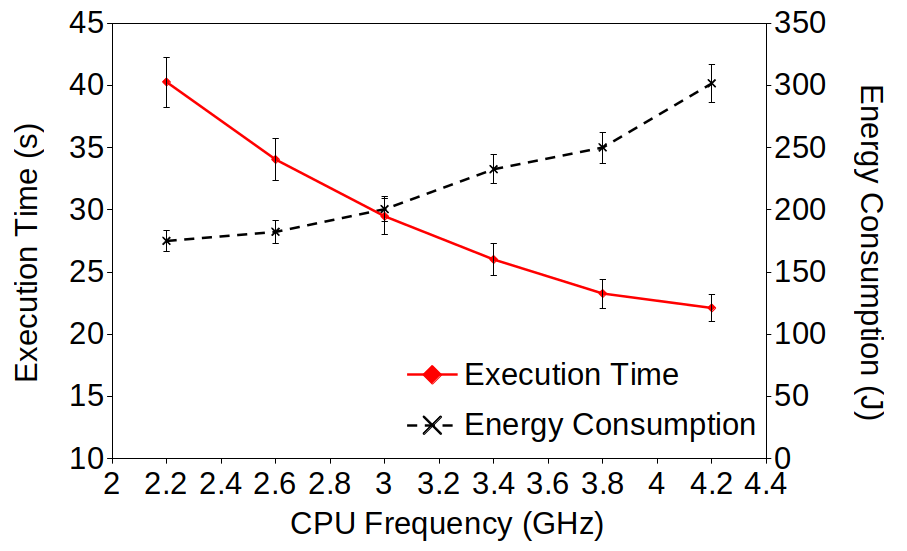}
    \caption{Execution Time vs Energy Consumption for a benchmarking workload}
    \label{fig:motivation}
\end{figure}

As mentioned above, modern CPUs enable the dynamic selection of their clock frequencies in order to save power during low-load conditions. Typically the operating system of the host machine is responsible to adjust that frequency through some pre-defined energy-saving policies. When running a CPU-intensive application, the application's performance is directly affected by the CPU frequency. If the CPU frequency is high, the application will be able to execute more instructions per second, leading to faster performance. On the other hand, if the CPU frequency is low, the application will be able to execute fewer instructions per second, leading to slower performance. Users expect different execution times from different applications. In Serverless Functions, for example, a user might expect a minimum processing throughput or a maximum response time for each request that arrives to the function. Response times of 250ms or 100ms both satisfy an SLO of the maximum response time of 300ms. 

As a motivation experiment, we demonstrate the performance behaviour of a CPU-intensive function running under different CPU frequencies while we also measure the consumed energy by the CPU. For our benchmark, we wrote a function in Go that performs 10,000,000 cycles of SHA256 on a string of 50 bytes once it is triggered by an external HTTP request. We packaged that function into a Docker Container and we run 4 instances in parallel on the same host running Ubuntu 20.04 LTS with an Intel(R) Core(TM) i7-7700 with 4 physical cores and Hyperthreading disabled. We tested frequencies from 2.2GHz up to 4.2GHz. We measured the consumed energy in Joules during the test with Powercap Util. We executed four requests in parallel, one for each instance, and we measured the response time of each request. Because the requests run in parallel all the process of all the requests took the same time. 

Thus the average execution time matches the execution time for each request. During the execution of the requests the \emph{CPU was constantly fully utilized}.

\begin{figure*}[t]
    \centering
    \includegraphics[width=0.8\linewidth]{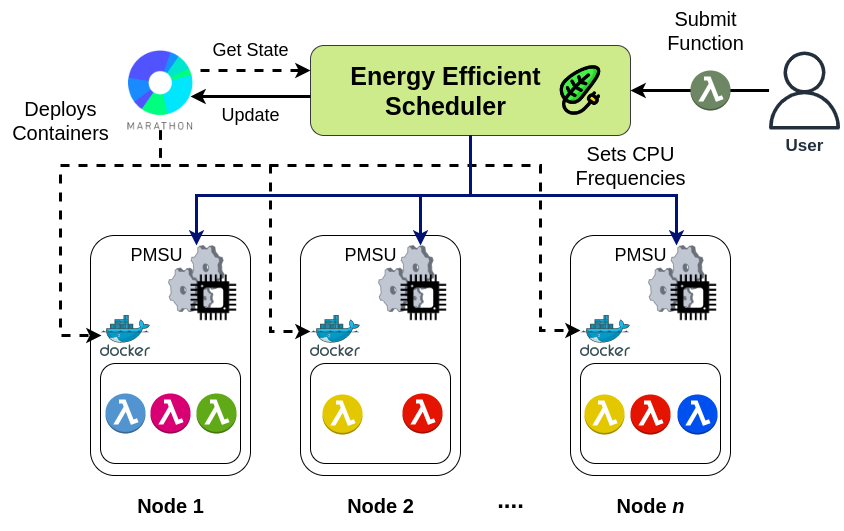}
    \caption{System Architecture}
    \label{fig:arch}
\end{figure*}

As we observe in figure \ref{fig:motivation}, as the CPU frequency increases the execution time decreases. For example, by reducing the CPU frequency from 4.0GHz to 3.6GHz we increase the execution time by 10\%, but we also reduce the energy consumption by 22.5\%. In cases where the response time is important, but not critical, this increase in latency will not affect the function SLOs. At the same time, Serverless Providers can benefit from reduced power consumption, which leads to lower operational costs and a reduced carbon footprint of the datacenter.

\section{Architecture}

Our system comprises two main components to operate the corresponding functionalities: (a) the Processor Management and Scheduling Utility and (b) the Energy Efficient Scheduler. Both components can be adapted in order to work in concert with any container orchestrator that allows the user to override the default scheduling policies or to explicitly select the container placement at the nodes.

\subsection{Processor Management and Scheduling Utility (PMSU)}
\label{pmsu}
The Processor Management and Scheduling Utility or PMSU for short, is a small system component that is responsible for the following functions: 

\begin{itemize}
    \item Adjust the clock frequency of the host CPU
    \item Report the current CPU clock frequency, temperature and power values
    \item Dispatch and schedule in real-time the containers to specific CPU cores
\end{itemize}

More specifically, the PMSU component receives frequency update requests via a custom TCP communication protocol. Upon receiving each frequency update request it returns a CONFIRM message back, while in the case that the request fails, it returns the error message and code.

The PMSU component is also responsible and periodically measures the instant power consumption of the CPU in Watts, and also the current CPU temperature. These measurements are then reported to the central Prometheus Monitoring server for further analysis and monitoring by the cluster operators. 

Finally, the PMSU component selects the CPU cores that each Docker Container uses at run-time. It intercepts the Container Creation event emitted from Docker and then by calling the Docker API, obtains the CPU resources that are allocated for that container. Once it selects a set of available CPU cores, it confines the container to a set of specific CPUs or cores. This approach resolves the Noisy Neighbor problem of multi-tenant cloud systems \cite{noisy} and the functions experience a more stable and predictable performance. Periodically, the PMSU transparently changes the CPU allocation of all containers in a round-robin scheme in order to avoid the creation of hot spots on the actual CPU die. 

\subsection{Energy Efficient Scheduler}

Another crucial component is the Energy Efficient Scheduler (EES). EES accepts scaling requests from the operators. Its objective is to exploit historic data, performance monitors and the provided energy-efficient scheduling policies, to determine the most appropriate set of worker nodes to allocate the function container replicas in order to reduce the total energy consumption of the entire cluster and satisfy the functions performance requirements. 

The EES component interacts with the container orchestrator via its API and orders it to place the containers in the selected nodes. 
Our scheduler is triggered whenever new incoming requests arrive at the cluster. 
If the newly incoming function can be scheduled, we place the container
in the appropriate node and it will be scheduled based on the frequency of the corresponding node. A function cannot be scheduled either in the case
that processing resources are not available at the nodes or in cases that 
processing resources are available but the addition of the new
function may cause existing functions to miss their deadlines.
In our implementation, we use Mesosphere Marathon\cite{marathon} on top of Apache Mesos\cite{mesos} as our container orchestrator, but it can easily adapt to any other container orchestrator that provides an API that can override the placement policies, such as for example, Kubernetes \cite{kubernetes}.

The EES stores scheduling parameters such as the function SLOs and cpu frequency demands, in the function deployment files that are stored in Marathon, by exploiting its labeling capabilities.

EES communicates with each cluster node PMSU via TCP connections, through which it can adjust the CPU clock frequency at each node. In our experimental evaluation we determined that EES is capable of sending around 1500 frequency update requests per second in our equipment.

\section{Energy Efficient Scheduler Methodology}

EES accepts scaling requests from users for both batch and stream processing jobs. When submitting a batch processing job, users are required to provide a deadline ($d$) and the total batch size ($bs$) of their job. On the other hand, for stream processing jobs, users need to specify the minimum desired throughput. EES comprises two components, the Scaling Component and the Scheduling Component. In Figure \ref{fig:flowchart}, we present a high-level flowchart illustrating the overall process of EES and providing a comprehensive overview of the key steps involved in the system's operation, highlighting the main stages and decision points.

\subsection{Scaling Component}
\label{scaler}
\begin{figure}[t]
    \centering
    \includegraphics[width=\linewidth]{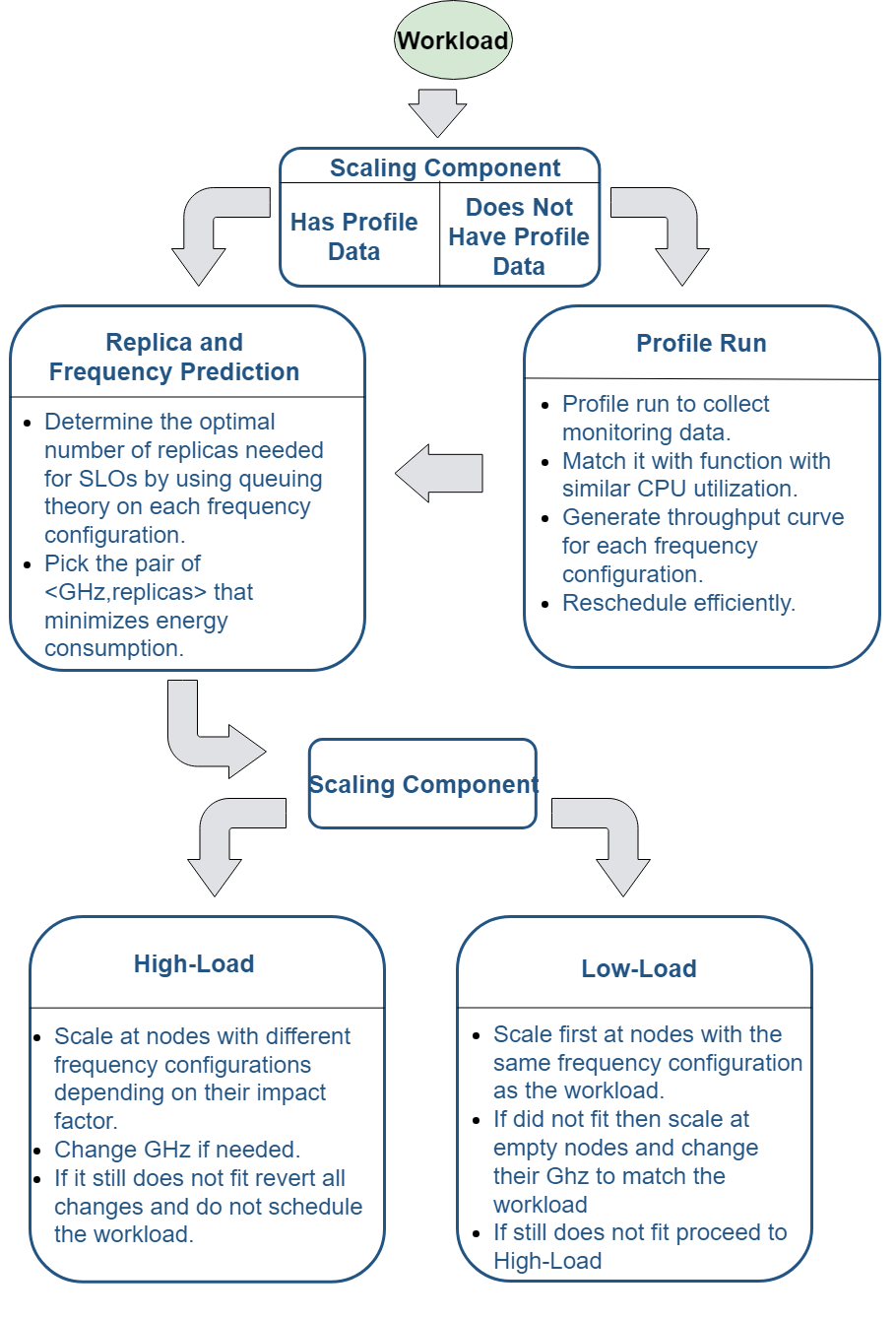}
    \caption{Energy Efficient Scheduler Flowchart}
    \label{fig:flowchart}
\end{figure}

The Scaling Component is responsible for determining the most suitable number of instances and their frequency configuration in order to satisfy (i) the user's Service Level Objectives (SLOs), and (ii) to minimize the overall energy consumption. EESc accomplishes this by leveraging performance monitoring information obtained from historical data of previous runs of the same job. When such data exists, the Scaling Component is capable of predicting the throughput and energy consumption of a single instance for each possible GHz configuration available. We do this by utilizing a queuing theoretical model to determine the optimal number of replicas necessary to fulfil the user's desired SLOs as shown on Fig.  \ref{fig:flowchart} (Replica and Frequency Prediction).

Let $\lambda$ denote the rate of request arrival. This rate is either directly specified by the user (in the case of stream processing jobs) or can be calculated from $\frac{bs}{d}$ (in the case of batch processing jobs). Under the assumption that the containers are homogeneous, we can conclude that all replicas $c$ will have the same average execution time ($avgExecTime$) for the same Ghz configuration. The $avgExecTime$ can be computed by exploiting the data of previous runs. Thus, we can compute the average service rate $\mu$ for each Ghz configuration $i$ as:

\begin{equation}
\label{mu}
    \mu_i= \frac{c_i}{avgExecTime_i}
\end{equation}

Assuming a Poisson distribution with $\lambda$ the rate for the incoming requests and also assuming that the services times are exponential, we can model this system using an M/M/c queuing system with $c$ function replicas to process incoming requests in parallel. The queuing analysis of an M/M/c system is widely documented in the literature. From this analysis, the stability of the utilization factor is considered a crucial metric that needs to be met. The utilization factor, denoted as $\rho$, represents the percentage of time the system is occupied with jobs. To calculate $\rho$ for each GHz configuration, the following formula is employed:

\begin{equation} 
\label{rho}
  \rho_i = \frac{\lambda}{\mu_i}
\end{equation}

In all queuing systems, the higher the average utilization factor, the longer the wait time for each job in the queue. Furthermore, when the average utilization factor exceeds 1, the queue size tends towards infinity, leading to an infinite average waiting time. As a rule of thumb, it is recommended to keep the utilization factor, denoted as $r$, below 80\% to avoid excessive waiting times and system instability.

To determine the optimal number of replicas $c_i$ for all possible frequency configurations in order to achieve $\lambda$, the Scaling Component solves Eq. (\ref{rho}) by combining it with Eq. (\ref{mu}). It then selects from all the available configurations $P$ the $<$GHz, $c_i>$ pair that minimizes the energy consumption $enrgCost_{i} * c_{i}$. The $enrgCost_{i}$ from each configuration is also obtained from historic data. The system then proceeds, by evaluating all possible configurations, to selecting the $<$GHz, $c_i>$ pair that minimizes the energy consumption:

\begin{equation} 
\label{argmin}
 \arg \min_{i\in P}(enrgCost_{i}*c_{i})
\end{equation}

The $enrgCost_{i}$ value for each configuration is also constructed from historical data.

Finally, if such information does not exist then the Scaling Component deploys the job with a random pair of $<$GHz,1$>$ and deploys it until it obtains some monitoring data Fig. \ref{fig:flowchart} (Profile Run). The monitoring data gathered from that profile run is the percentage of the CPU utilized and the throughput of the function for that specific frequency. EES matches it with the function with the closest CPU utilization. As we observed, functions with similar CPU utilization have similar energy consumption. The throughput will follow a similar curve to the throughput curve of the similar function, displaced in the Y-axis by  $$ r=\frac{throughput_{gathered}}{throughput_{known}} $$ 

\begin{figure}[h]
    \centering
    \includegraphics[width=\linewidth]{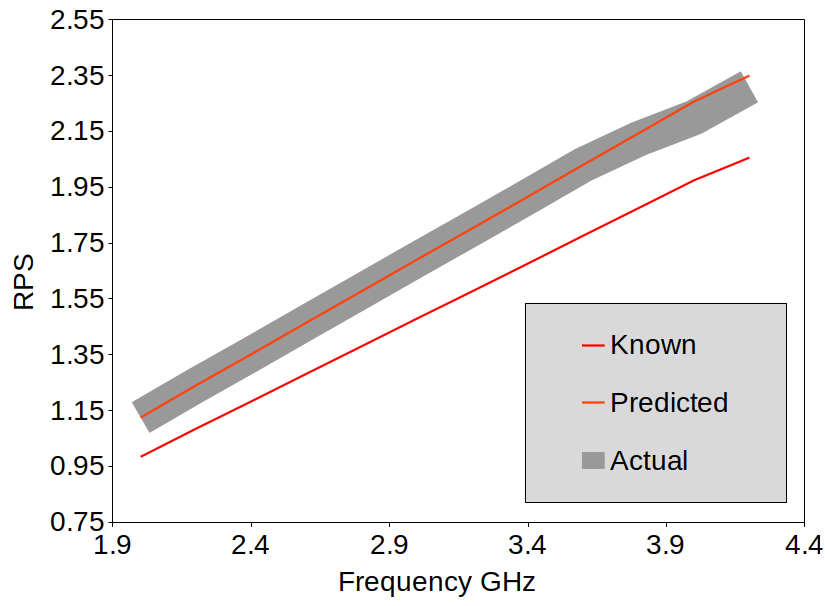}
    \caption{Predicting an unknown function throughput (in Request Per Second - RPS) based on a known function with similar CPU utilization}
    \label{fig:prediction}
\end{figure}

In Fig. \ref{fig:prediction} we demonstrate how EES, given a known function, can estimate the throughput on each frequency configuration of a function with an unknown profile but with \emph{similar CPU utilization} with a low deviation from the actual throughput. In this figure, we used 'Sha256 1' from Table \ref{table:workloads} to estimate the throughput of 'Sha256 3' which has different CPU allocation and different input size. As we can see, EES predicted the throughput with high accuracy.

For jobs without historic data meeting the SLOs are not guaranteed while the actual profiling is done by the EES and for this reason, we propose the user to perform a small profile run before submitting the actual workload.

\subsection{Scheduling Component:} The Scheduling Component is responsible for deploying jobs within the cluster. It follows the early binding scheduling model which means that jobs are immediately deployed for execution on the most appropriate available node. This approach is preferred because it can provide reduced startup latency which is crucial for real-time serverless applications with strict deadlines that have to be met. However, it is important to note that early binding scheduling does come with some drawbacks. One disadvantage is that there is no queuing stage, which means the system cannot reorder function scheduling to achieve more optimal allocations on the nodes. Despite its drawbacks in that regard, we believe that this trade-off is acceptable for the FaaS model. For our scheduling, we propose a straightforward but quite effective hybrid greedy load balancing scheduling model which changes depending on the load. 

\textbf{Low-Load:} During periods of low load i.e. when there are nodes with no running tasks or there are nodes with running tasks but whose set frequency matches the job's frequency the Scheduling Component deploys them using the following greedy approach Fig. \ref{fig:flowchart} (Low-Load). First, it starts from the nodes whose frequency matches the job's frequency. It then proceeds to assign function instances to available cores on these nodes, starting with the nodes that have the fewest available cores. If there are no more cores available then it picks arbitrarily from the empty nodes and after setting their frequency to match the function's frequency it starts placing replicas to them. If there are no more available empty machines the algorithm proceeds to the high-load load balancing model.

\textbf{High-Load:} During periods of high load i.e. when all nodes are hosting jobs and the nodes with available cores do not match the job's frequency, the Scheduling Component employs a different greedy approach for scheduling Fig. \ref{fig:flowchart} (High-Load). For each node with available cores, our algorithm assesses the impact of deploying the job there. If the machine is currently running at a higher frequency than the job's frequency, deploying the job there would result in it running at a higher frequency than the one that we have identified as the most appropriate. The impact in this case would be the increased energy cost we would incur. On the other hand, if the machine is running at a lower frequency, the Scheduling Component would have to increase the node's frequency, where now the impact is the increased energy cost of the functions already running on that node, compared to their previous lower frequency. It then sorts the nodes depending on that impact factor and starts deploying replicas at the nodes. If there are insufficient available cores prior to job scheduling, the job will not be scheduled, and the user will be notified accordingly. Although it is possible to deploy multiple jobs in the same core we opted against it because it causes unpredictable results for the job. 

Following this greedy approach, we (i) maximize the probability that all jobs will meet their SLOs, and (ii) in cases of high-load we try to lower the energy impact where the optimal frequency for maximum energy reduction cannot be guaranteed. We achieve these goals without resorting to rescheduling, which could lead to better scheduling but at a cost of higher job delays and unpredictable outcomes.

Finally, when a job is completed, it is removed from the nodes. If the finished job had the highest frequency among the running jobs, the Scheduling Component adjusts the frequency to the second highest. In the case there are no other jobs running, it sets the frequency to the lowest available option.

\section{Implementation}

We have implemented our techniques in a prototype system in order to experimentally evaluate the working and benefits of our approach. For our serverless system, we use OpenFaas Function templates for building our application functions as Docker Images. We use Mesosphere Marathon 1.5 on top of Apache Mesos 1.9 as our container Orchestrator. A high-throughput HTTP reverse proxy written in Java acts as the Gateway to the function containers. It listens to container start / stop events received from Marathon and then it can automatically proxy the incoming function requests to the available function replicas. It also logs execution statistics for each function and reports them to Prometheus through the Prometheus Java client.

\begin{table*}[!htb]
     \caption{Workloads Description}
\centering
\begin{tabular}{|l|l|l|l|l|} 
\hline
Function & Language & CPU       & Network & Description                                        \\ 
\hline
Cars     & Python   & High      & High    & Cars detection in a given image                   \\
Sha256   & Go       & High      & Low     & Performs N loops of~SHA256~over a given string     \\
Linpack  & Python   & Very High & Low     & Solves a dense system of linear equations of size \emph{N} \\
Pdf      & Python   & Low       & Medium  & Generates a PDF file~by~combining text with image  \\
\hline
\end{tabular}
   \label{table:types}

\end{table*}

We have also implemented the Energy Efficient Scheduler (EES) which includes all the described scheduling policies. The Scheduler is written in Java and uses the Marathon API for deploying containers to our cluster nodes. It overrides the default scheduling policies of Marathon by creating custom application deployments which exploit node placement constraints. In this way, it can instantiate an arbitrary number of function replicas across all cluster available nodes.

The PMSU is written in Java and exploits the Docker Http API in order to interact with the Docker Engine for setting the CPU affinity (cpuset) of the running containers. It receives Frequency Update requests from the EES and uses the Debian package cpupower for setting the CPU frequency of the node. It also uses the Debian package Rapl in order to get the current CPU energy consumption and lm-sensors for monitoring the current CPU temperature. It uses the Java Prometheus client to report in real-time the power and temperature measurements. 

For the Monitoring service, we use Prometheus \cite{prometheus} and Grafana\cite{grafana} for generating real-time graphs for the various parameters that we monitor. Prometheus is a time-series database with proven performance and querying capabilities. 

\section{Evaluation}

In this section we present the setup and methodology we used in our experimental evaluation.

\subsection{Experimental Setup}

We evaluated our Energy Efficient Scheduling techniques in our  local computer cluster which comprises twelve nodes in total. Seven of these nodes are running Ubuntu 20.04 LTS with Linux Kernel 5.15 where each node is equipped with an Intel(R) Core(TM) i7-7700 CPU with 4 physical cores, base frequency at 3.60 GHz, Max Turbo Frequency at 4.2 GHz and a TDP of 65 Watts. Each of these nodes has 16 GBs of RAM. Thus, we had 28 physical cores for deploying and executing the application functions. 
Similarly to \cite{kafes}, we deactivated Hyper-Threading to obtain more predictable results. 
In addition, we used five separate nodes for running the EES, the Mesos and Marathon Master and Zookeeper, the Prometheus and Grafana services, the Gateway and the load generator that is used for injecting the user workloads. All nodes are interconnected through a 1 Gbps network.

\begin{figure*}[h]
\minipage{0.49\textwidth}
  \includegraphics[width=\linewidth]{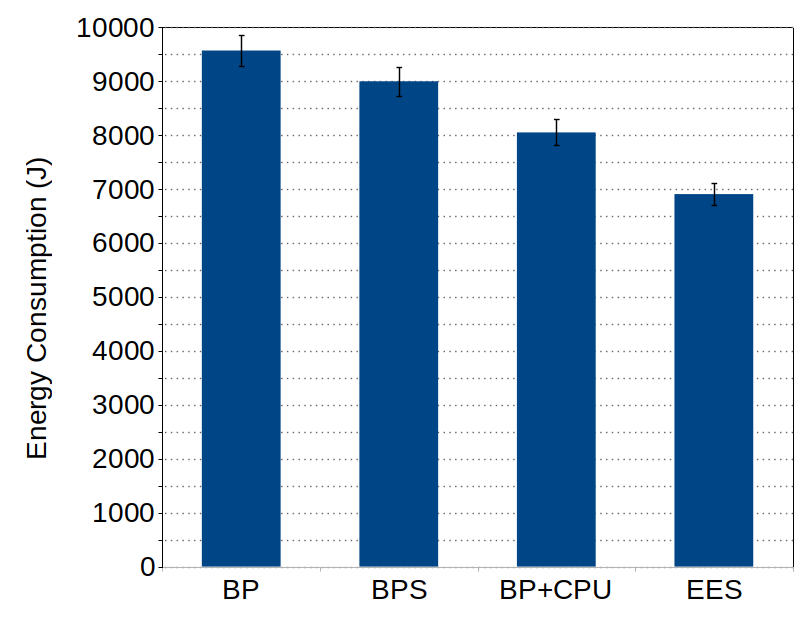}
  \caption{Total energy consumption in Joules}\label{fig:energy}
\endminipage\hfill
\minipage{0.49\textwidth}
  \includegraphics[width=\linewidth]{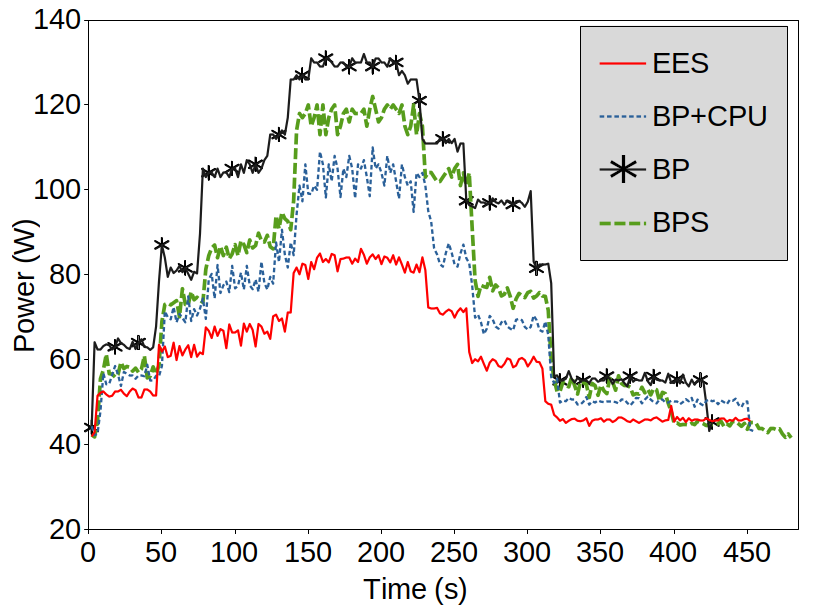}
  \caption{Cluster power consumption at runtime with the different techniques}\label{fig:watts}
\endminipage\hfill
        \label{fig:result-util}
\end{figure*}

\begin{figure*}[h]
\minipage{0.49\textwidth}
  \includegraphics[width=\linewidth]{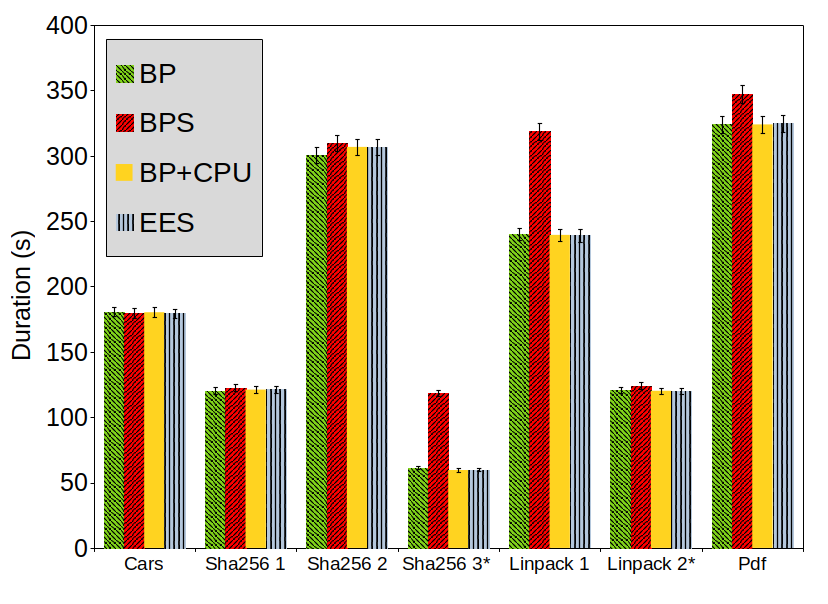}
  \caption{Comparison of EES with State of the Art techniques in terms of Workload Duration for different workloads}\label{fig:duration}
\endminipage\hfill
\minipage{0.49\textwidth}
  \includegraphics[width=\linewidth]{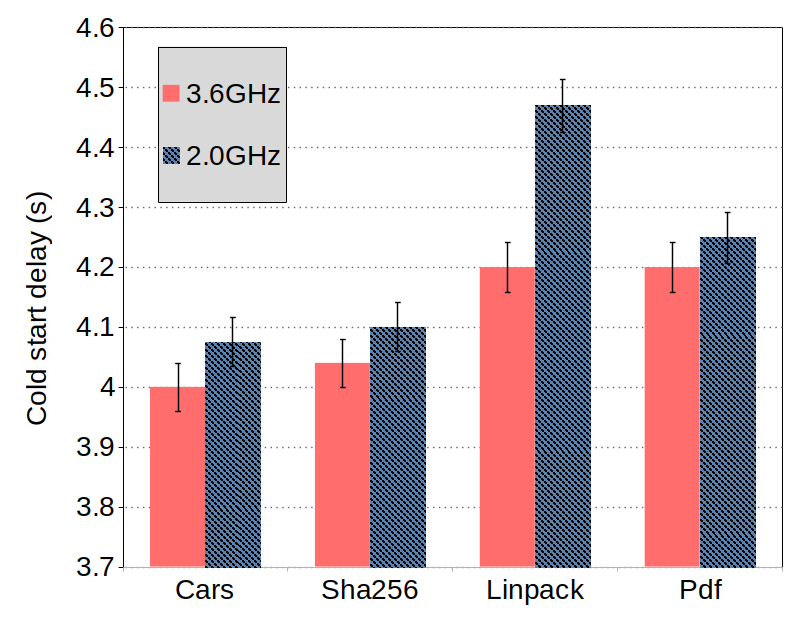}
  \caption{Cold-start delay with minimum and maximum frequency configuration for different workloads }\label{fig:coldstart}
\endminipage\hfill
        \label{fig:result-util}
\end{figure*}

\begin{figure}[h]
    \centering
    \includegraphics[width=\linewidth]{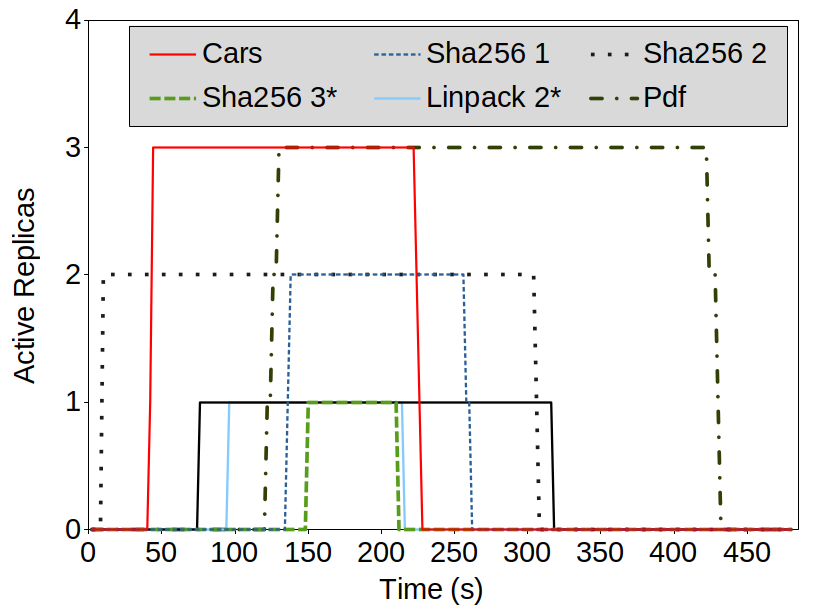}
 \caption{Active Replicas For Each Workload with EES}\label{fig:replicas}
\end{figure}

\begin{figure}[h]
    \centering
    \includegraphics[width=\linewidth]{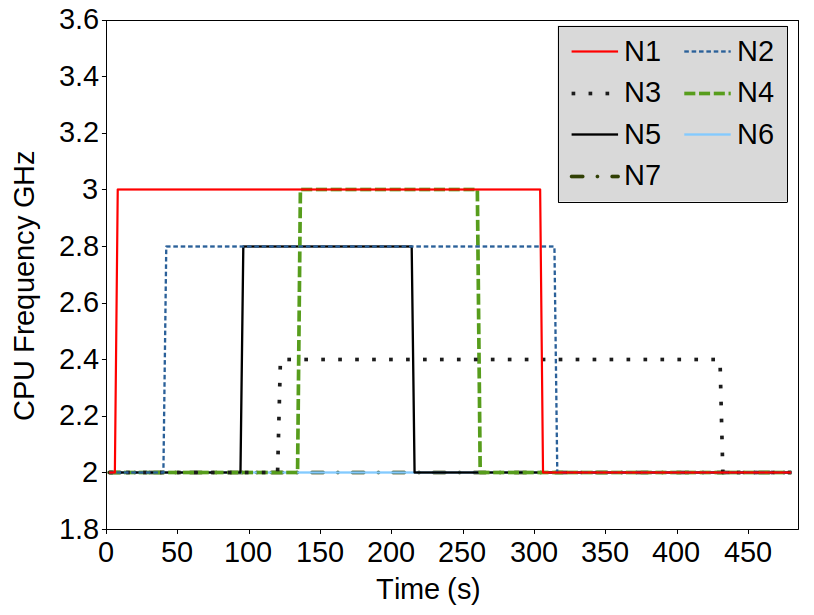}
 \caption{Frequency selection for each node by EES}\label{fig:frequency}
\end{figure}

\subsection{Evaluation Methodology}

In order to provide a detailed experimental evaluation of our proposed methods, we designed and conducted the following set of experiments. Our aim in the experiments was to provide insights into how the following aspects of a serverless FaaS system are affected by our proposed methodology. 
\begin{itemize}
    \item Energy consumption
    \item Cold-start
    \item Performance
\end{itemize}

\begin{table*}[!htb]
\centering
     \caption{Workload Allocations}
\begin{tabular}{|l|c|c|c|c|c|c|} 
\hline
Workload       & \multicolumn{1}{l|}{BP replicas} & \multicolumn{1}{l|}{BPS replicas} & \multicolumn{1}{l|}{BP+CPU replicas} & \multicolumn{1}{l|}{EES replicas} & \multicolumn{1}{l|}{Baselines Freq. Range configuration} & \multicolumn{1}{l|}{EES Freq. configuration}  \\ 
\hline
Car Detection  & 4 replicas                       & 4 replicas                        & 3 replicas                            & 3 replicas                        & 2.0-3.6 GHz                                          & 2.8 GHz                                     \\
Sha256 1       & 2 replicas                       & 2 replicas                        & 2 replicas                            & 2 replicas                        & 2.0-3.6 GHz                                          & 3 GHz                                       \\
Sha256 2       & 3 replicas                       & 2 replicas                        & 3 replicas                            & 2 replicas                        & 2.0-3.6 GHz                                          & 3 GHz                                       \\
Sha256 3*       & 1 replica                        & 1 replica                         & 1 replica                             & 1 replica                         & 2.0-3.6 GHz                                          & 3 GHz                                       \\
Linpack 1      & 1 replica                        & 1 replica                         & 1 replica                             & 1 replica                         & 2.0-3.6 GHz                                          & 2.8 GHz                                     \\
Linpack 2*      & 6 replicas                       & 6 replicas                        & 1 replica                             & 1 replica                         & 2.0-3.6 GHz                                          & 2.8 GHz                                     \\
Pdf Generation & 3 replicas                       & 3 replicas                        & 3 replicas                            & 3 replicas                        & 2.0-3.6 GHz                                          & 2.4 GHz                                     \\
\hline
\end{tabular}
  \label{table:replicas}
\end{table*}

We evaluated our EES scheduler against three other techniques. The first technique that we compared is the Baseline Performance (BP). In this technique, we used the default scheduling policies of Mesosphere Marathon and Apache Mesos for the container placement of the functions, and the Linux Performance Power governor in the cluster nodes for power management. For the second technique, Baseline Powersave (BPS), we used again the same scheduling policies from Marathon, but we used the Powersave power governor of Linux in the cluster nodes. The Powersave governor sets the CPU clock frequency to the lowest setting. In our case, 2.0GHz was set as the lowest, however during the experiments we observed that the frequency jumped to over 3.0GHZ which implies that either the Kernel is not fully compatible with the CPU or the CPU overrides this setting on its own. Finally, we compared our policy against BP, but this time we used cpuset parameter of the Docker containers in order to confine the execution of each function container to a set of specified cores of the processor as described in the section \ref{pmsu}. This technique is defined as BP+CPU in the experimental figures.

For the three Baseline approaches we implemented a \emph{Request Per Second} auto-scaler which is a standard scaling technique used in many serverless systems such as OpenFaas and KNative \cite{knative}. The auto-scaler periodically observes the incoming traffic for each function and calculates the required replicas as:
$$replicas = ready\:replicas \cdot \frac{mean\:load\:per\:replica }{target\:load\:per\:replica }$$

For our EES scheduler, we calculated the replicas according to the method in Section \ref{scaler}. The EES available configurations start from 3.6Ghz and can go down to 2GHz in fixed intervals of 200MHz

In parallel, we tested how the CPU frequency affects the cold start of the functions. We tested the 4 different functions with the minimum and maximum CPU frequencies that our scheduler uses.

\subsection{Workloads}

For our workloads, we used two standard benchmarks with high CPU demands (Sha256, Linpack)\cite{functionBench}, one with high I/O writes in the storage system Minio\cite{minio} (Pdf) and one with high CPU demand and high I/O read writes to Minio (Cars). The functions used as benchmarks are shown in the following Table \ref{table:types}. \emph{Sha256} and \emph{Linpack} are the two benchmark functions that aim to load the CPU to its limit. Sha256 performs N loops of SHA256 cryptographic hash function over a given string. The number N is given as a parameter at the request. For our experiments, the N value is 1000000. Linpack is a widely used benchmark workload, also used by Top500 \cite{top500}, that solves a dense system of linear equations of size N. Similarly with SHA256 that N is given as a parameter at the request. We created two Linpack workloads (i) a light one where N is equal to 100 and (ii) a large one with N equal to 1000. \emph{Pdf} is a high throughput function that generates a Pdf document by combining text and an image which are read from Minio. Finally, \emph{Cars}, uses the YOLOv5s \cite{yolo} object detection model to find the cars on an image that is read from Minio and then it stores the position of their bounding box as a text file back to it. 

We deployed these functions with different CPU allocations and parameters in our cluster as shown in Table \ref{table:workloads}, these were injected simultaneously in the cluster. For each deployment, we selected a request rate from Microsoft’s Azure traces \cite{serverlessInTheWild,traces,azure} and we used Hey \cite{hey} to generate traffic. The inter-arrival time between the workloads is given by a Poisson distribution with $\lambda = 0.05$ and a fixed random seed for repeatability across the experiments. For the workloads that are indicated with "*" in their name, we did not have profile runs and used the appropriate technique that is described in section \ref{scaler}. For the rest of the workloads, we used profiles acquired from historic data.

\section{Evaluation Results}

\subsection{Energy consumption}

The main goal of our EES scheduler is to minimize the total power consumption while maintaining the required performance. From Figure \ref{fig:energy} we observe that EES is from 14\% up to 28\% 
superior to its competitors in terms of  
energy savings. In particular, EES achieves 28\% energy savings compared to the BP approach, which is the standard frequency setting and scheduling technique used in cluster nodes. EES has significant benefits both in terms of energy savings and performance over BPS, the default powersaving mode of the baseline technique (as we also discuss later in Figure \ref{fig:duration}). 
In Figure \ref{fig:watts} we depict the cluster power consumption at runtime with the different
techniques. As the figure shows, the lowest power consumption was achieved with EES at all times, compared to all other methods. Note though, that, Table \ref{table:replicas} illustrates that EES and BP+CPU deployed fewer replicas compared to BP and BPS for the same request rate which leads to lower total energy consumption compared to BP and BPS. Furthermore, due to the lower frequency configuration, EES achieves even lower energy consumption compared to BP+CPU. When using the CPUs in low power, the datacenters can benefit in more ways that the direct reduced energy consumption. For example, by keeping the CPUs in low power, the datacenter consumes less energy for operating cooling units, because the CPUs run cooler. 

\subsection{Performance}
As we observe in Fig. \ref{fig:duration} our EES scheduler achieved the same or better performance compared to the three other techniques. This denotes that our methodology achieves its dual objective, {\it i.e.,} minimize the cluster energy consumption while meeting the performance goals ({\it i..e,} SLOs) of the scheduled functions. We also observe that the Linux Powersave governor leads to unpredictable behaviour in the majority of the workloads.

\subsection{Cold-start}
In terms of cold-start delay, in Fig. \ref{fig:coldstart} we can observe that the average cold-start delay slightly increased with the lower CPU frequency with the exception of Linpack which increased by 260ms. We have to note here, that, our nodes use HDD drives instead of SSDs or NVMe drives which reduce the cold start of the containers by multiple orders of magnitude. As a result, we can conclude that by reducing the CPU frequency, our scheduler adds minimal extra delay to the cold start of the function instances and also that as expected, higher CPU frequencies result in slightly lower cold start delays.

\subsection{Runtime System Operation}
Figures \ref{fig:replicas} and \ref{fig:frequency}
illustrate the operation of our system at run-time for the different workloads. 
In Figure \ref{fig:replicas} we show that the different workloads 
require different numbers of replicas in the deployed functions. 
Furthermore, in Figure \ref{fig:frequency} we show the 
frequency selection for each node in our cluster. We observe that 
using our approach all nodes run at lower frequencies than the base frequency of the CPU (3.6 GHz). Furthermore, two of the nodes (N6, N7) are idle, which indicates that the EES achieved better resource utilization than all its competitors.

\subsection{Evaluation Discussion}
As we can observe from the results above, EES has important benefits as it manages to achieve very similar performance with the baseline approaches in terms of throughput, while at the same time consuming less energy due to the lower frequency configurations of the nodes' CPUs. The observed results validate our hypothesis that optimizing scheduling strategies can lead to reduced energy consumption. 

Furthermore, from our experimental results we can confidently say that implementing Energy-Efficient Scheduling can maintain the workloads on the same overall performance while minimizing energy consumption. By leveraging the lower frequency configurations of the nodes' CPUs, EES effectively balances the trade-off between throughput and energy efficiency. 

\begin{table}
\centering
     \caption{Workloads Resources}
\begin{tabular}{|l|c|l|c|c|c|c|} 
\hline
Workload       & CPU Cores & Memory  & Rate Per Second   \\ 
\hline
Car Detection  & ~1        & 1024 MB & 3.9 Rps                   \\
Sha256 1       & 1         & 125 MB  & 3.2~Rps                     \\
Sha256 2       & 1         & 125 MB  & 3.3~Rps                        \\
Sha256 3*      & 0.5       & 125 MB  & 1~Rps                          \\
Linpack 1      & 1         & 1024 MB & 1~Rps                               \\
Linpack 2*      & 0.5       & 1024 MB & 14~Rps                           \\
Pdf Generation & 0.5       & 256 MB  & 138Rps                         \\
\hline
\end{tabular}
  \label{table:workloads}
\end{table}

\section{Related Work}

There has been significant prior work in terms of scheduling batch and stream processing workloads \cite{hybrid} \cite{tomaras2} trying to satisfy applications’
performance demands. In our previous work, \cite{amesos} we employed queuing theory to optimize Serverless Streaming Pipelines. Yu et al. in \cite{faasrank} used Reinforcement Learning (RL) to learn automatically the scheduling policies through experience in clusters running serverless functions for minimizing the average invocation execution time. Authors in \cite{ensure} propose a function-level scheduler and resource manager for serverless computing. It dynamically classifies and manages function requests to minimize infrastructure costs while meeting performance requirements. In their work, Kaffes et al. \cite{kafes} investigate scheduling techniques for serverless systems. They introduce a scheduler that takes into account factors such as cost, load, and locality to minimize the occurrence of cold starts, in contrast to load-based policies, while ensuring high-performance levels. All of these works focus only on efficient scheduling but they do not consider possible energy savings.

Jeong et al. in \cite{faas-energy} propose an energy-efficient service scheduling algorithm for federated edge cloud (FEC) environments. Their approach aims to minimize the total energy consumption and reduce QoS violations. The algorithm optimizes service placement and path selection by considering actual traffic requirements, leading to improved energy efficiency and reduced service violation rates compared to existing approaches. Maroulis et al. in ExpREsS \cite{express}, \cite{stathisJournal} propose  a scheduler designed to minimize energy consumption while meeting performance requirements by leveraging time-series prediction models for energy usage and execution times and applying a DVFS technique for Apache Spark batch and stream workloads. Although their technique is effective it is not applicable to serverless environments due to the multi-tenancy of the hosts where an action for energy minimization for one function can affect the performance of another since they share the same physical CPU. Other approaches such as MicroFaaS \cite{microfaas} exploit low-powered edge ARM-based single-board computers to run functions. Due to the low power demand, they achieve lower energy consumption. Although these approaches are interesting, they lack computing power and they cannot be used for multi-tenant systems or resource-demanding workloads.

\section{Conclusion}
In this work we have studied the problem of energy-efficient scheduling in multi-tenant serverless cloud systems. We have advanced the state-of-the-art in several ways: (1) showed that the performance and power demand of applications executing in serverless systems expressed as a set of stateless functions can be affected by different cpu frequencies,
(2) provided a Processor Management and Scheduling Utility (PMSU) component
that can dynamically adjust the clock frequency of the host CPU 
and obtain CPU clock frequency, temperature and power values 
at run-time which are exploited by the scheduling component when
making allocation decisions,  and
(3) proposed a novel Energy Efficient Scheduler, EES that determines the most appropriate set of worker nodes to allocate the function container replicas in order to reduce the total energy consumption of the entire cluster and satisfy the functions performance requirements.
Our experimental results indicate a clear improvement in the system’s dual objective of meeting performance goals and minimizing the cluster's energy consumption when our methodology is used, outperforming current state-of-the-art techniques.

\section*{Acknowledgement}
The research work was supported by the Hellenic Foundation for Research and Innovation (HFRI) under the 3rd Call for HFRI PhD Fellowships (Fellowship Number: 6812), by the European Union through the EU ICT-48 2020 project TAILOR (No. 952215), the H2020 AutoFair project (No. 101070568) and the Horizon Europe CoDiet project (No. 101084642).
\bibliographystyle{IEEEtran}
\bibliography{refs}
\end{document}